\documentclass[aps,showpacs,10pt]{revtex4}
\usepackage{graphicx}
\usepackage{epsfig}
\usepackage{bm}
\usepackage{amsmath}
\usepackage{amssymb}
\usepackage{graphics}
\begin{document}
\title{An application of transverse-momentum-dependent evolution 
equations in QCD}
\author{Federico A. Ceccopieri}
\email{ceccopieri@fis.unipr.it}
\affiliation{Dipartimento di Fisica, Universit\`a di Parma,\\
Viale delle Scienze, Campus Sud, 43100 Parma, Italy}
\author{Luca Trentadue}
\email{luca.trentadue@cern.ch}
\affiliation{Dipartimento di Fisica, Universit\'a di Parma,
INFN Gruppo Collegato di Parma, Viale delle Scienze, Campus Sud, 
43100 Parma, Italy}
\begin{abstract}
The properties and behaviour of the solutions of 
the recently obtained $k_t$-dependent evolution equations
are investigated. When used to reproduce transverse momentum spectra 
of hadrons in Semi-Inclusive DIS, an encouraging agreement with data 
is found. The present analysis also supports at the phenomenological level 
the factorization properties of the Semi-Inclusive DIS cross-sections in terms of 
$k_t$-dependent distributions. 
Further improvements and possible developments of the proposed evolution equations are
envisaged. 

\end{abstract}

\maketitle
\section{Introduction}
\label{sec:intro}

In a standard perturbative QCD approach to semi-inclusive processes and in particular to 
Semi-Inclusive Deep Inelastic Scattering (SIDIS), 
factorization theorems~\cite{fproof} allow to extract soft hadronic wave functions 
from high energy reactions data. Such non-perturbative process-independent distributions 
obey QCD renormalization group equations~\cite{DGLAP}.
In presence of a hard scale, set by the virtuality of the exchanged boson 
in a Deep Inelastic event, standard parton and fragmentation distributions predict, 
together with the corresponding process-dependent coefficient 
functions~\cite{SIDIS_start,SIDIS_Cij}, the semi-inclusive cross-sections.
These distributions, basic ingredients in almost nowaday QCD-calculations,
are well suited for studying full inclusive process, such as 
Deep Inelastic lepton-hadron Scattering or Drell-Yan process in hadronic collisions.
In the recent past however it has became increasingly clear the less inclusive distributions, 
either space-like or time-like, are necessary to deal with a variety of 
semi-inclusive processes. In particular $k_t$-dependent distributions
acquired particular relevance and a great activity has been registered recently
in this research field \cite{KMR}.
In the SIDIS case, for istance, final state hadrons are expected to have 
a sizeable transverse momentum due to both intrinsic motion of partons into 
hadrons~\cite{Chan} and to the radiative process off the struck parton 
line~\cite{SIDIS_Cij,Chay}. Unfortunately transverse momentum is 
usually integrated over, loosing part of the information which is contained in the experimental 
cross-sections.
For these reasons it would be highly desiderable to have the evolution equations for 
these extended $k_t$-dependent distributions. Such evolution equations were first 
proposed in the time-like case in Ref.~\cite{BCM,NT} and then recently 
extended in the space-like domain in Ref.~\cite{our_work}.
In order to have a complete description of the semi-inclusive cross-sections
in terms of the transverse momentum, such a generalization was also performed 
in the target fragmentation region by introducing properly modified~\cite{our_work} fracture
functions~\cite{Trentadue_Veneziano}. 
The basic idea behind the $k_t$-dependent evolution equations can be summarized as follows. 
Let us consider parton emissions off a active, space-like, parton line. 
In the collinear limit, at each branching,  the generated transverse momentum is negligible.
In this limit however $k_t$-ordered diagrams can be shown to give 
leading logarithmic enhancements to the cross-sections.
Since such contributions can be resummed by DGLAP evolution equations~\cite{DGLAP}, 
at the end of the radiative process, the interacting parton could possibly
have an appreciable transverse momentum. As a result, $k_t$-dependent evolution equations 
therefore depend, in addition to standard longitudinal momentum fraction, also on transverse
degree of freedom. When solutions to the evolution equations are used to reproduce 
the SIDIS transverse momentum spectrum, the predictions smooth interpolate from 
small to large transverse momenta, this being a signature of well known DIS 
scaling violations in semi-inclusive process. 

The aim of this work is to offer a preliminar phenomenological study 
of $k_t$-dependent evolution equations and to compare it with 
available hadron production data in DIS current fragmentation region. All the predictions 
are given by a handful of phenomenological assumptions. However such predictions
are not the result of a fit to data, and thus strengthen our confidence 
in the general framework offered in Ref.~\cite{our_work}.    

\section{Transverse momentum dependent evolution equations}
\begin{figure}[ht]
\begin{center}
\includegraphics[width=5.4cm,height=2.8cm,angle=0]{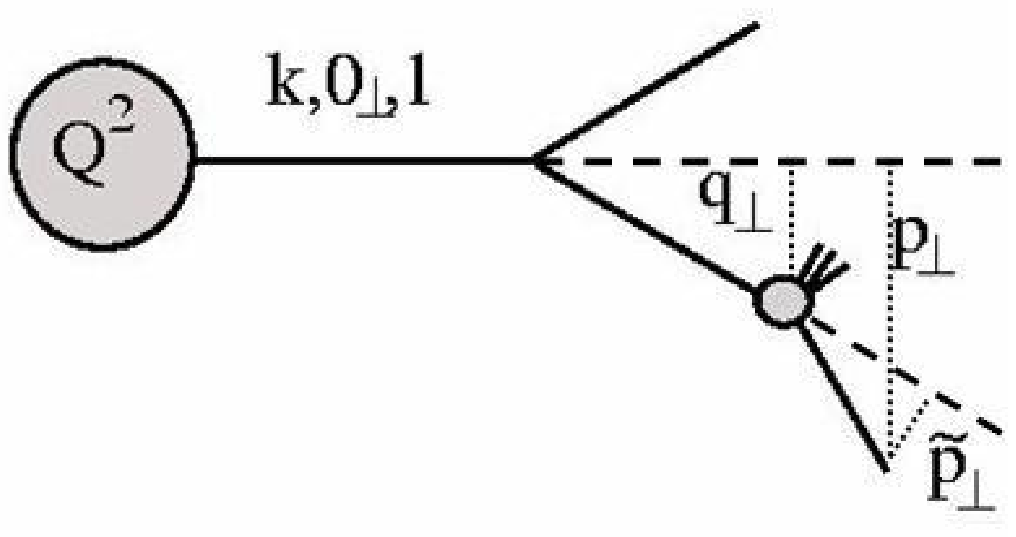}
\includegraphics[width=5cm,height=3cm,angle=0]{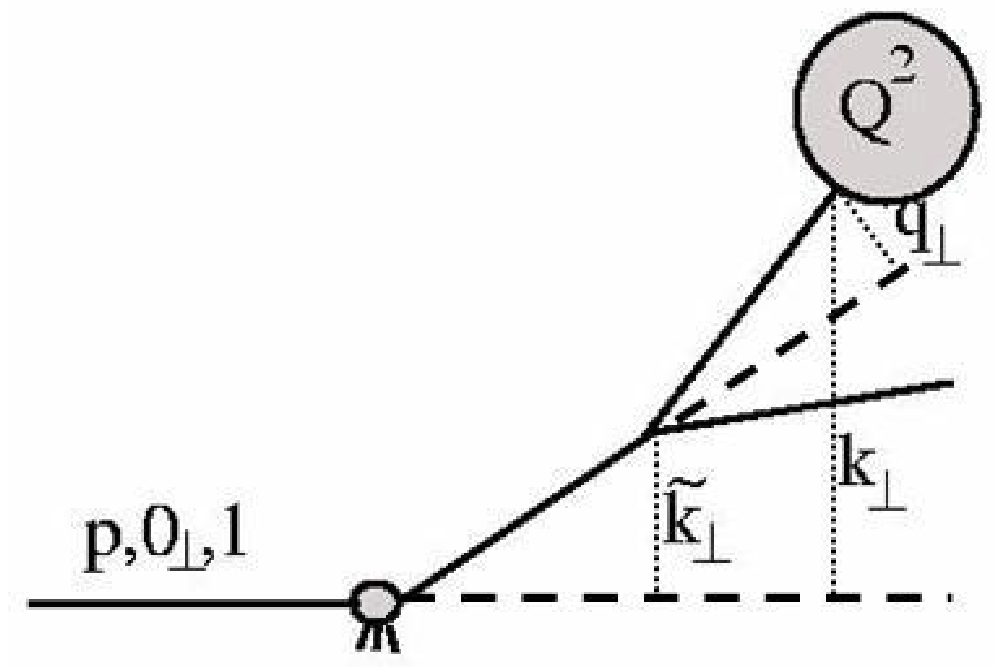}
\caption{Boost of transverse momenta.  
Left panel: a time-like off-shell parton generated in a hard process, the grey blob, 
emits a daughter parton and acquires a transverse momentum $\bm{q}_{\perp}$
relative to its intial direction. The small blob symbolizes the iteration of such emissions. 
Right panel: the analogue as before in the space-like case.}
\label{boost} 
\end{center}
\end{figure}

Ordinary QCD evolution equations at leading logarithm accuracy (LLA)
resum terms of the type $\alpha_s^n \log^n (Q^2/\mu^2_{F})$ originating
from quasi-collinear partons emission configurations, 
where $\mu^2_{F}$ represents the factorization scale.
Leading contributions are obtained when 
the virtualities of the partons in the ladder are strongly ordered.
At each branching, the emitting parton thus acquires a transverse momentum 
relative to its initial direction. 
The radiative transverse momentum 
can be taken into account through transverse-momentum-dependent evolution equations, 
which in the time-like case read~\cite{BCM}:
\begin{equation}
\label{dglap_TMD_time}
Q^2 \frac{\partial \mathcal{D}_{i}^{h}(z_h,Q^2,\bm{p_{\perp}})}{\partial Q^2}=
\frac{\alpha_s(Q^2)}{2\pi}\int_{z_h}^1 \frac{du}{u} 
P_{ij}(u,\alpha_s(Q^2)) \int \frac{d^2 \bm{q_{\perp}}}{\pi}\,\delta(\,u(1-u)~Q^2-q^2_{\perp})\,
\mathcal{D}_{j}^{h}\Big(\frac{z_h}{u},Q^2,\bm{p_{\perp}}-\frac{z_h}{u} \bm{q_{\perp}} \Big).
\end{equation}
Fragmentation functions $\mathcal{D}_{i}^{h}(z_h,Q^2,\bm{p_{\perp}})$ 
of eq.~(\ref{dglap_TMD_time}) give
the probability to find, at a given scale $Q^2$, a hadron $h$ 
with longitudinal momentum fraction $z_h$ and transverse momentum $\bm{p}_{\perp}$ 
relative to the parent parton $i$.
$P_{ij}(u)$ are the time-like splitting functions which, at least at LL accuracy, 
can be interpreted as the probabilities
to find a parton of type $i$ inside a parton of type $j$ and are 
expressed as a power series of the strong running coupling,
$P_{ij}(u)=\sum_{n=0}\alpha_s^{n}(Q^2)P_{ij}^{(n)}(u)$.
The order $n$ of the expansion of the splitting function matrix $P_{ij}(u)$ 
actually sets the accuracy of the evolution equations. 
The radiative transverse momentum square $q_{\perp}^2$ at each branching 
satisfies the invariant mass constraint  $q^2_{\perp}=u\,(1-u)\,Q^2$ \,.
The transverse arguments of $\mathcal{D}_{i}^{h}(z_h,Q^2,\bm{p}_{\perp})$ on r.h.s. of eq.~(\ref{dglap_TMD_time})
are derived taking into account the Lorentz boost of transverse momenta
from the emitted parton reference frame to the emitting parton one, see the left panel of 
Fig.~(\ref{boost}).
\newline The unintegrated distributions fulfil the normalization:
\begin{equation}
\label{timelike_norm}
\int d^2 \bm{p}_{\perp} \mathcal{D}_{i}^{h}(z_h,Q^2,\bm{p}_{\perp})=\mathcal{D}_{i}^{h}(z_h,Q^2)\,.
\end{equation}
This property garantees that we can recover ordinary integrated distributions from 
unintegrated ones. The opposite statement however is not valid since eq.~(\ref{dglap_TMD_time}) 
contains new physical information. 
In analogy to the time-like case we consider now a initial state parton $p$
in a incoming proton $P$ which undergoes a hard collision, 
the reference frame being aligned along the incoming proton axis.
We thus generalize eq.~(\ref{dglap_TMD_time}) to the space-like case \cite{our_work}:
\begin{equation}
\label{dglap_TMD_space}
Q^2 \frac{\partial \mathcal{F}_{P}^{i}(x_B,Q^2,\bm{k_{\perp}})}{\partial Q^2}
=\frac{\alpha_s(Q^2)}{2\pi}\int_{x_B}^1 \frac{du}{u^3} 
P_{ji}(u,\alpha_s(Q^2)) \int \frac{d^2 \bm{q_{\perp}}}{\pi}\,\delta(\,(1-u)Q^2-q^2_{\perp})
\,\mathcal{F}_{P}^{j}\Big(\frac{x_B}{u},Q^2, \frac{\bm{k}_{\perp}-\bm{q}_{\perp}}{u} \Big)\,.
\end{equation}
Parton distribution functions $\mathcal{F}_{P}^{i}(x_B,Q^2,\bm{k_{\perp}})$ 
in eq.~(\ref{dglap_TMD_space}) give
the probability to find, at a given scale $Q^2$, a parton  $i$ 
with longitudinal momentum fraction $x_B$ and transverse momentum $\bm{k}_{\perp}$ 
relative to the parent hadron, see the right panel of Fig.~(\ref{boost}).
The unintegrated distributions fulfil a condition
analogous to the one in eq.~(\ref{timelike_norm}), \textsl{i.e.} :
\begin{equation}
\label{spacelike_norm}
\int d^2 \bm{k}_{\perp} \mathcal{F}_{P}^{i}(x_B,Q^2,\bm{k}_{\perp})=
\mathcal{F}_{P}^{i}(x_B,Q^2)\,.
\end{equation}
We note that the inclusion of transverse momentum does not affect longitudinal 
degrees of freedom since partons  
always degrade their fractional momenta in the perturbative branching process.

The approach can also be extended in the target fragmentation region 
of semi-inclusive DIS \cite{our_work}
by introducing a $k_t$-dependent version of fracture functions~\cite{Trentadue_Veneziano}.
The corresponding evolution equations for 
$\mathcal{M}^{i}_{P,h}(x,\bm{k_{\perp}},z,\bm{p_{\perp}},Q^2)$ can be obtained
\cite{our_work}. The factorization properties of these distributions however 
has not been proven yet, at variance with the current fragmentation case whose 
factorization in terms of $k_t$-dependent distributions has been proven in Ref.~\cite{Ji}.

\section{Phenomenology in the current fragmentation region}
\label{sec:Phenomenology} 
The $k_t$-dependent evolution equations, eq.~(\ref{dglap_TMD_time}) 
and eq.~(\ref{dglap_TMD_space}), 
are solved  by means of a finite difference method in the  $(2n_f+1)$-dimensional 
space of quarks, antiquarks and gluons. As appropriate for a leading logarithmic calculations,
we set splitting functions to their lowest order expansion.
In this preliminar analysis we simulate light flavours only while   
heavy flavours are accounted for in only as virtual contributions 
in the LL running coupling constant, $\alpha_s(Q^2)$.
Convolutions on transverse and longitudinal 
variables in eq.~(\ref{dglap_TMD_time}) and eq.~(\ref{dglap_TMD_space}) 
are numerically performed 
on a bidimensional $(x, k_{\perp}^2)$ grid. To achieve a faster convergence and minimize the size of the grid, non-linear spacing both in $x$ and in $k_{\perp}^2$ have been adopted.
At each $Q^2$-iteration, the normalization conditions, 
eq.~(\ref{timelike_norm}) and eq.~(\ref{spacelike_norm}),
are checked to reproduce ordinary longitudinal distributions within a given accuracy.
\begin{figure}[t]
\begin{center}
\includegraphics[width=18cm,height=6cm,angle=0]{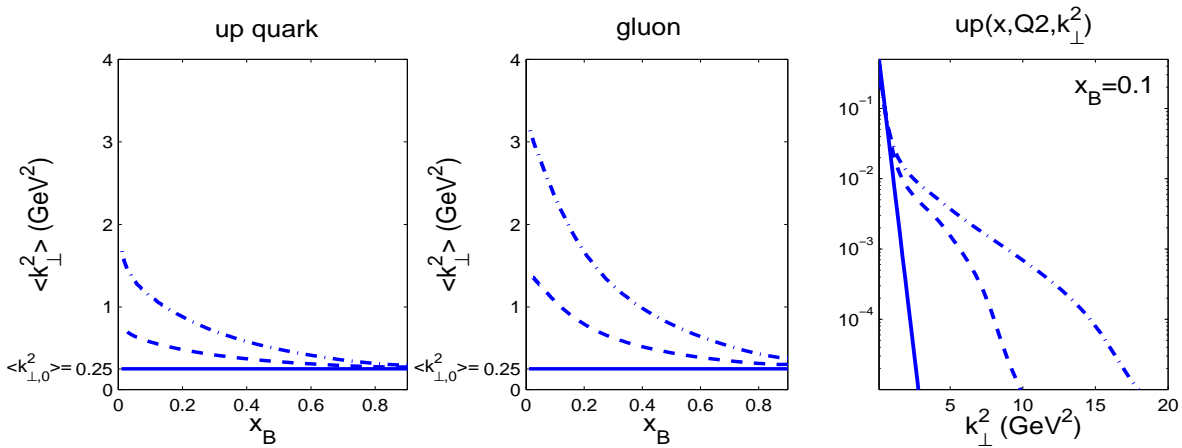}
\caption{ Space-like evolution. Left and middle panel: average transverse momentum $<k_{\perp}^2>$ 
generated in the evolution of the up-quark and gluon for three different scales:
$Q_0^2=5$  Ge$V^2$ ($-$), $Q^2=10$ Ge$V^2$ ($--$) and $Q^2=20$ Ge$V^2$ ($-\cdot$)\,.
Right panel: the transverse spectrum of the up quark at fixed $x_B$ for three different scales 
as before. The solid line is the gaussian initial condition. The evolved distributions 
show a $1/(k_{\perp}^2)^{\gamma}$ dependence.}
\label{fig2}
\end{center}
\end{figure}
As in the longitudinal case, $k_t$-dependent distributions at a
scale $Q^2>Q_0^2$ are calculable if one provides a non-perturbative input 
density at some arbitrary scale $Q_0^2$.
In the following we assume the simplest, physically motivated ansatz, \textsl{i.e.}
a longitudinal parton distribution function $F_P^i(x_B,Q_0^2)$~\cite{MRST2001} 
or fragmentation functions $D^{h}_{i}(z_h,Q_0^2)$~\cite{Kretzer} times a gaussian transverse factor, 
motivated by the Fermi motion of partons in hadrons~\cite{Chan}: 
\begin{equation}
\label{ansatz}
\mathcal{F}_P^i(x_B,Q_0^2,\bm{k}_{\perp})= F_P^i(x_B,Q_0^2)
\,\frac{e^{\frac{-k_{\perp}^2}{<k_{\perp,0}^2>}}}{\pi<k_{\perp,0}^2>}\,, 
\;\;\;\;\; 
\mathcal{D}^{h}_i(z_h,Q_0^2,\bm{k}_{\perp})=D_i^{h}(z_h,Q_0^2) 
\,\frac{e^{\frac{-p_{\perp}^2}{<p_{\perp,0}^2>}}}{\pi<p_{\perp,0}^2>} \,\,\,
\;\;\; i=q,\bar{q},g\,.
\end{equation}
Before comparing to data, we would like to draw some general properties of the evolution and discuss
the stiffness of the initial conditions, eq.~(\ref{ansatz}). We focus on the space-like case 
and set the width $<k_{\perp,0}^2>$ to a testing value of $0.25$ Ge$V^2$ both for quarks and gluons. 
The evolution then is performed from the initial scale $Q_0^2=$ $5$ Ge$V^2$ to 
$Q^2=$ $20$ Ge$V^2$, see Fig.~\ref{fig2}. 
In order to reduce the number of parameters, we assume a flavour-independent
value for the average transverse momentum $<k_{\perp,0}^2>$. Such hypothesis is indeed too crude
in the quark valence region.
Furthemore, as it appears in Fig.~(\ref{fig2}), the evolution generates 
a $x_B$-dependent amount of averaged transverse momentum,
behaving like
\begin{equation}
\label{x-dep}
 <k_{\perp}^2>=<k_{\perp,0}^2>\, x_B^{\gamma}, \;\; \gamma \leq 0 \,,
\end{equation}
even starting from a $x_B$-independent distribution, eq.~(\ref{ansatz}).
This behaviour is expected since the arguments of $k_t$-dependent distributions
in the right hand side of eq.~(\ref{dglap_TMD_time}) and eq.~(\ref{dglap_TMD_space})  
mix, as a result of transverse boost, longitudinal and transverse degree of freedom. We have checked that the factorized form 
of eq.~(\ref{ansatz}) is not preserved under evolution and deviation from a guassian form
into broader $k_{\perp}^2$-distributions, especially for the gluon, are observed. 
In the rightmost panel of Fig.~(\ref{fig2}) is clearly visible how the evolution 
turns the the gaussian transverse factor at the initial scale into a inverse power-like distributions 
in $k_{\perp}^2$ at the final scale. 
It is also visible in the same plot a de-population effect in the $k_{\perp}^2 \simeq Q^2$ region according 
to strong ordering recipe built-in the evolution equations.
From above arguments and since the factorization scale $Q_0^2$, at which 
we suppose eq.~(\ref{ansatz}) to be valid, is arbitrary we conclude
that a more refined analysis could use initial condition with a $x_B$-dependent transverse factor.  
We note also that the solutions do not show any growth of  $<k_{\perp}^2>$ in the large 
$x_B$ limit. In the soft limit the $k_t$-dependent evolution equations can be shown 
to diagonalize in impact parameter phase \cite{Parisi} by a joint Fourier-Mellin transform \cite{BCM}. 
As a result soft gluon resummation technique can be applied to leading and next-to-leading 
logarithmic accuracy~\cite{KT,CSS,Nadolsky}.  
Attaining these limitations in mind we compare in the following the outcome 
of $k_t$-dependent evolution equations 
with charged hadron production data in the DIS current fragmentation region.
In this case we are supported by a factorization theorem 
and the semi-inclusive cross-sections can be shown to factorize in terms 
of $k_t$-dependent distributions~\cite{Ji}.
With leading logarithmic accuracy the cross-sections reads
\begin{equation} 
\label{kt_sidis_fact}
\frac{d^5 \sigma}{dx_B \,dQ^2\, dz_h\, dQ^2\, d^2 \bm{P}_{h\perp}}\propto\sum_{i=q,\,\bar{q}}
e_i^2\int d^2\bm{k}_{\perp} d^2\bm{p}_{\perp} \, \delta^{(2)} 
(z_h \bm{k}_{\perp}+\bm{p}_{\perp} - \bm{P}_{h\perp}) 
\mathcal{F}_{P}^{i}(x_B,Q^2,\bm{k}_{\perp},) 
\;\mathcal{D}^{h}_{i}(z_h,Q^2,\bm{p}_{\perp})\;,
\end{equation}
while the soft factor, present in the original factorization formula in Ref.~\cite{Ji}, 
is dropped for phenomenological purposes.
The standard SIDIS variables are defined as $x_B=Q^{2}/(2 P \cdot q)$
and $z_h=(P\cdot P_h) / (P\cdot q)$ where $P,~P_h,~q$ are respectively the four momenta 
of the incoming proton, outgoing hadron and virtual boson.   
At lowest order, the process-dependent coefficient function is omitted and set to unity.
Factorizations scales are set to $\mu^2_{F}=\mu^2_{D}=Q^2$ 
and large logarithmic ratios of the type $\log(\mu^2_{F,D}/Q^2)$ 
occuring in the perturbative calculations
are moved in $k_t$-dependent distributions are then resummed by evolution equations.
We compare our predictions with data of Refs.~\cite{EMC80,EMC91}. 
\begin{figure}[t]
\begin{center}
\includegraphics[width=15cm,height=6cm,angle=0]{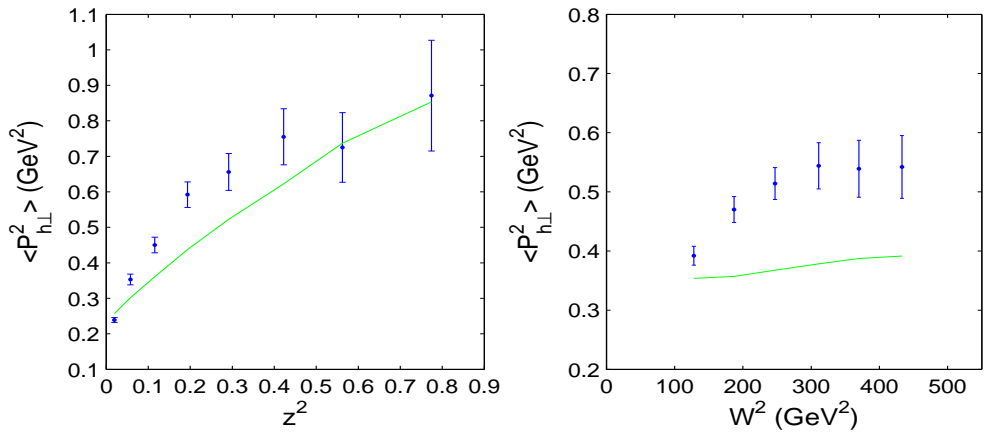}
\caption{ Left panel: average transverse momentum $<P_{h\perp}^2>$ versus $z^2$,
$100<W^2<340$ Ge$V^2$, $Q^2>5$ Ge$V^2$, against  
predictions (solid line). Right panel:  
average transverse momentum $<P_{h\perp}^2>$ versus $W^2$ for $0.2<z<1.0$, $Q^2>5$ Ge$V^2$, 
against predictions (solid line). Data from Ref.~\cite{EMC80}\,.}
\label{fig3}
\end{center}
\end{figure}
These data sets are differential in the variable of interest 
and cover a broad kinematical region, where DGLAP dynamics is supposed to be valid. 
We note that the theoretical predictions which reproduce the data in the original Refs.~\cite{EMC80,EMC91}
are based on QCD-calculations of Ref.~\cite{Altarelli_predictions}.
Light flavours average transverse momenta and factorization scale are then set to  
\begin{equation}
<k_{\perp, q, \bar{q}}^2> = 0.25 \;\makebox{GeV}^2\,, \;\; 
<p_{\perp, D_q, D_{\bar{q}}}^2> = 0.20 \;\makebox{GeV}^2\,,
 \;\; Q_0^2=5\;\makebox{GeV}^2\,,
\end{equation}
for distribution and fragmentation functions respectively, according to Ref.~\cite{Anselmino}.
The parameters in Ref.~\cite{Anselmino} are obtained by a fitting procedure 
to the low-$P_{h\perp}^2$ differential cross-sections 
of Ref.~\cite{EMC91} using the same initial condition as given in eq.~(\ref{ansatz}). 
We note that gluons in eq.~(\ref{kt_sidis_fact}) are absent since do not directly couple 
with the virtual boson 
but enter indirectly the cross-sections due to quark-gluon 
mixing in the evolution equations.
Gluon widths are however essentially unknown and for this reason, in this preliminar analysis, 
we set them equal to light flavours parameters. In order to verify that this choice does not  
affect the presented results,  
we have checked that a $20\%$ variation of gluon widths does not alter significantly the predictions 
in the kinematical region of Refs.~\cite{EMC80,EMC91}, the overall effect being a 
slightly slope variation of the large-$P_{h\perp}$ tail in Fig.~(\ref{fig4}).
The role of gluon and its transverse spectrum is however of special interest especially
in HERA and LHC kinematics, and thus certainly deserves a separated study. 
We require both the time-like and space-like evolved $k_t$-dependent distributions to satisfy, 
both for quarks and gluons, the normalization condition, 
eq.~(\ref{timelike_norm}) and (\ref{spacelike_norm}), in the kinematical 
range of data~\cite{EMC91,EMC80} with an accuracy set to $10\%$:
\begin{equation}
\int d^2 \bm{k}_{\perp} \mathcal{F}_i(x,k_{\perp},Q^2)\Big|_{\mbox{\tiny{EMC}}}
=F_i(x,Q^2)\Big|_{\mbox{\tiny{EMC}}}\,.
\end{equation}
The accuracy however could by increased properly thickening the simulation grid.
In Fig.~(\ref{fig3}) we show the average transverse momentum  $<P_{h\perp}^2>$ compared to the predictions
of eq.~(\ref{kt_sidis_fact}) properly normalized to the relevant inclusive cross-sections. 
In the left panel a rise of $<P_{h\perp}^2>$ with $z^2$ is observed. 
Essentially this dependence is guided by the $\delta^{(2)}$-function in eq.~(\ref{kt_sidis_fact})
which leads to the expectation $<P_{h\perp}^2>=<p_{\perp}^2>+z^2<k_{\perp}^2>$ \,. 
The slope of the data is roughly reproduced.
On the right panel of Fig.~({\ref{fig3}}) the $<P_{h\perp}^2>$ results 
obtained from eq.~(\ref{kt_sidis_fact}) as a function of 
$W^2=Q^2 (1-x_B)/x_B$ are compared to data. 
The $<P_{h\perp}^2>$ spectrum shows a clear logarithmic dependence on $W^2$ and 
the predictions far underestimate the measured average transverse momentum. 
As can be seen in Ref.~\cite{EMC80}, the measured dependence 
of $<P_{h\perp}^2>$ on $Q^2$ is very mild, while the one on $x_B$ is steeper and of a kind shown 
in Fig.~(\ref{fig2}). The $k_t$-dependent evolution equations take care of the former while 
probably only a $x_B$-dependent correction in the transverse factor in eq.~(\ref{ansatz})
could solve the latter.
The charged hadron production $P_{h\perp}^2$-differential 
cross-sections, properly integrated in the relevant $z$ and $W^2$ bins and normalized to the inclusive total cross-sections, 
is shown in Fig.~(\ref{fig4}). 
\begin{figure}[t]
\begin{center}
\includegraphics[width=17cm,height=16cm,angle=0]{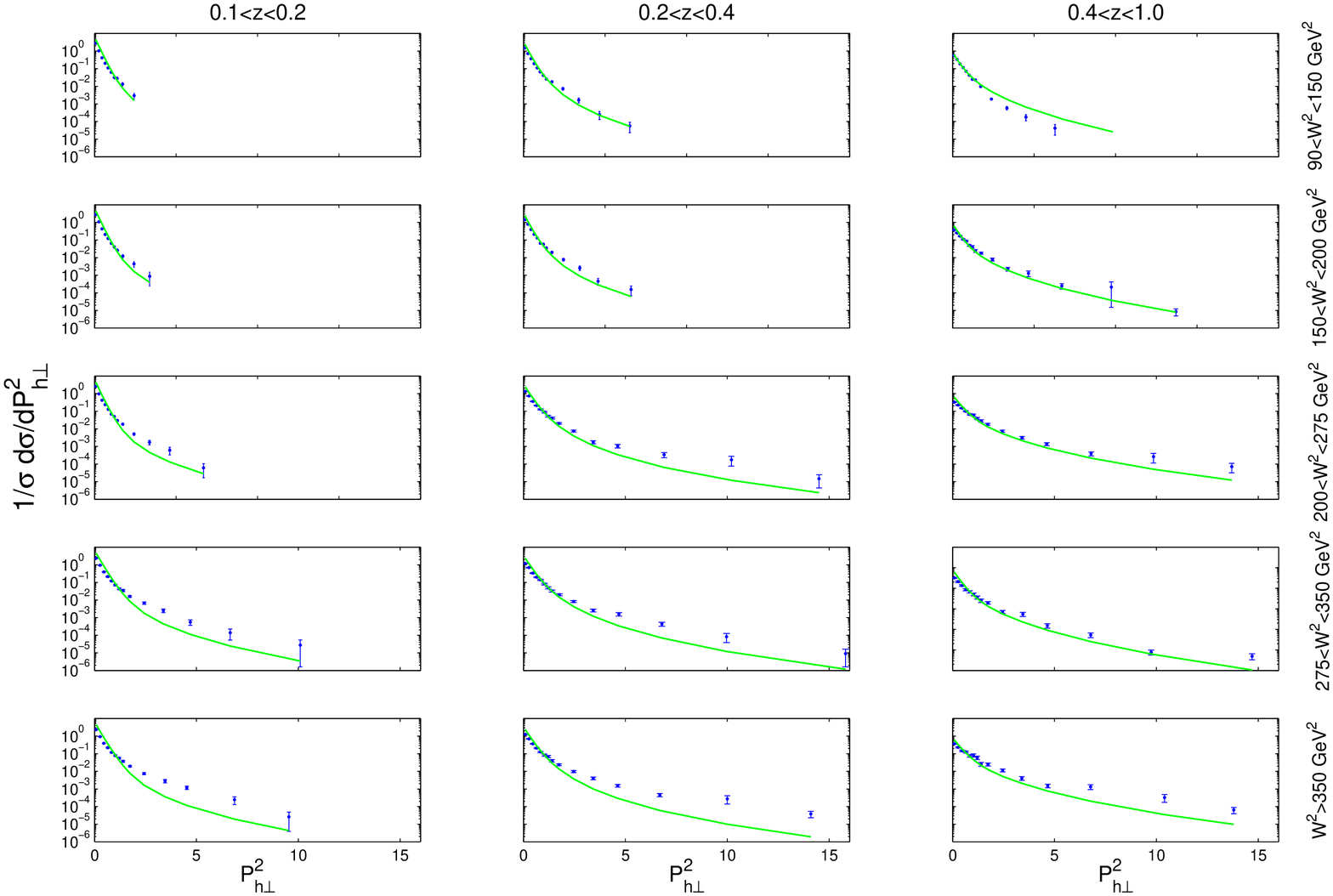}
\caption{ The normalized semi-inclusive cross-sections in bins of $z$ and $W^2$ 
for charged hadron production in the current fragmentation region. Data from Ref.~\cite{EMC91}\..} 
\label{fig4}
\end{center}
\end{figure}
The main effect of evolution equations is of modifying the 
sharp tail of the guassian distributions at $Q_0^2$, 
see Ref.~\cite{Anselmino}, into a broader transverse power-like distributions. 
At fixed $W^2$, a progressive 
broadening of the spectrum, according to the left panel of Fig.~(\ref{fig3}), is observed. 
At fixed $z$ instead, the predictions fall more distant from data 
as long as $W^2$ increases, according with Fig.~(\ref{fig3}).
The overall agreement looks however encouraging since we have not performed any fit to the data, 
apart from fixing the transverse widths as already discussed.   
At high $W^2$, in the low-$P_{h\perp}^2$ part of the spectrum, deviations in slope 
beetwen data and predictions are visible, signalating again the inadeguacy of a 
$x_B$-independent width.
The underestimation of the transverse spectrum at high $P_{h\perp}^2$ indicates instead  
that large angle parton emissions from fixed order matrix element are needed. 
We conclude that both a more accurate choice of the initial condition 
and the inclusion of next-to-leading corrections 
will lead thus to a better agreement of the predicted cross-sections with data.

\section{Conclusions}
\label{sec:conc}
In this work, by using the factorization theorem of Ref.~\cite{Ji}, 
the charged hadron production cross-sections in the current fragmentation region has been computed
within leading logarithmic approximation by using $k_t$-dependent evolution equations.
The obtained $k_t$-dependent distributions, due to resummation of soft and collinear parton emissions,
reproduce the high $P_{h\perp}^2$ tail of tranverse spectra
and smoothly interpolate from the low to the high $P_{h\perp}^2$ regime 
without using any matching procedure between the two regions. 
A reasonable description of the data is obtained by only 
using default width values as proposed in Ref.~\cite{Anselmino}. 
This validates our approach as proposed in Ref.~\cite{our_work}.
The impact of the intial conditions, eq.~(\ref{ansatz}), is investigated and 
arguments for a $x_B$-dependent transverse factor are given, along 
with hints suggesting the need of higher order corrections. 
In this work we do not emphasize gluon dynamics.  
This subject however is a central one, especially for HERA and LHC kinematics \cite{KMS,JMY}, 
and we deserve it for a separate study.

We wish to conclude by listing two possible promising applications of the presented formalism. 
The $k_t$-dependent evolution equations could be tested in Drell-Yan pair production 
cross-sections differential in the transverse momentum of the lepton pair. 
The present formalism could find interesting applications to polarized reactions 
and could be particularly fruitful, for istance, in the case of transversity ditributions~\cite{h1}.

\section{Acknowledgements}
F.A.C. acknowledge fruitful discussions at the Trento Workshop `Transverse momentum, 
spin, and position distributions of partons in hadrons' with J.~C.~Collins, W.~Vogelsang and M.~Diehl.

\end{document}